\documentclass[%
 reprint,
groupedaddress,
 amsmath,amssymb,
 aps,
]{revtex4-2}

\usepackage{graphicx}
\usepackage{dcolumn}
\usepackage{bm}
\usepackage{color}
\usepackage{amsmath}
\usepackage{hyperref}

\begin{document}

\title{Quantum Griffiths singularity in three-dimensional MoTiN\\
 superconducting films}

\author{Zi-Xiao Wang}
\affiliation{Tianjin Key Laboratory of Low Dimensional Materials Physics and
Preparing Technology, Department of Physics, Tianjin University, Tianjin 300354,
China}
\author{Tian-Yu Jing}
\affiliation{Tianjin Key Laboratory of Low Dimensional Materials Physics and
Preparing Technology, Department of Physics, Tianjin University, Tianjin 300354,
China}
\author{Zi-Yan Han}
\affiliation{Tianjin Key Laboratory of Low Dimensional Materials Physics and
Preparing Technology, Department of Physics, Tianjin University, Tianjin 300354,
China}
\author{Kuang-Hong Gao}
\affiliation{Tianjin Key Laboratory of Low Dimensional Materials Physics and
Preparing Technology, Department of Physics, Tianjin University, Tianjin 300354,
China}
\author{Song-Ci Li}
\affiliation{Tianjin Key Laboratory of Low Dimensional Materials Physics and
Preparing Technology, Department of Physics, Tianjin University, Tianjin 300354,
China}
\author{Zhi-Qing Li}
\email[Corresponding author, e-mail: ]{zhiqingli@tju.edu.cn}
\affiliation{Tianjin Key Laboratory of Low Dimensional Materials Physics and
Preparing Technology, Department of Physics, Tianjin University, Tianjin 300354,
China}
\date{\today}

\begin{abstract}
Quantum Griffiths singularity (QGS) has been experimentally observed in a range of two-dimensional (2D) superconducting systems. Although it is theoretically suggested that the QGS also exists in three-dimensional (3D) superconductors, there is almost no experimental support to the theoretical prediction. In the present paper, we observe the occurrence of QGS in a series of $\sim$80-nm-thick Mo$_{0.8}$Ti$_{0.2}$N$_x$ ($0.84 \lesssim x \lesssim 1.12$) superconducting films near the field-driven superconductor-metal transition (SMT). These films have a NaCl structure and are 3D with respect to the superconductivity. For each film, the low-temperature magnetoresistance isotherms, measured at magnetic fields being perpendicular or parallel to the film plane, do not cross at a single point but in a wide region. The dynamical critical exponents $z\nu_{\perp}$ (for perpendicular field) and $z\nu_{\parallel}$ (for parallel field) obtained by analyzing the related magnetoresistance isotherms increase with decreasing temperature and tend to diverge as $T\rightarrow 0$\,K. In addition, the effective resistivity data for the perpendicular and parallel field in the vicinity of the SMTs both obey an activated scaling based on the random transverse-field Ising model. The QGS in the 3D Mo$_{0.8}$Ti$_{0.2}$N$_x$ superconducting films originates from the slow dynamics of the rare regions in these systems. We also fabricate a $\sim$80-nm-thick (Mo$_{0.8}$Ti$_{0.2}$)$_2$N$_{1.06}$ superconducting film with face-centered cubic structure at low nitrogen partial pressure. It is found that the low-temperature magnetoresistance isotherms for the perpendicular (parallel) field cross at a single point and the resistivity data for the perpendicular (parallel) field in the vicinity of the field-induced SMT obey the power-law scaling deduced from the dirty-boson model. Our results provide unambiguous experimental evidence for the existence of QGS in 3D superconductors.
\end{abstract}

\maketitle


\section{Introduction}\label{secI}
Quantum Griffiths singularity (QGS) has been discovered in a range of two-dimensional (2D) superconducting systems and attracted great attention over the past decade~\cite{SciXing, prbShen, NcSaito, NLXing}. Conceptually, the QGS refers to the phenomenon that certain thermodynamic observables of a system are singular not just at criticality but in a finite region in the vicinity of the quantum critical point. As for the 2D superconducting systems, the main characteristic of QGS is that the low-temperature magnetoresistance isotherms do not cross at a point near the critical field but at multiple points, and the dynamical critical exponent $z\nu$ obtained at each crossing point diverges as $T\rightarrow0$\,K~\cite{SciXing, prbShen, NcSaito, NLXing}. Experimentally, the existence of QGS was initially reported in three-dimensional (3D) magnetic materials, including 3D heavy fermion systems and $f$-electron ferromagnetic alloys~\cite{prldeAndrade, prlCastroNeto, SciSteppke, prlUbaidKassis}. Until 2015, the QGS was observed in 3-monolayer Ga films, providing experimental evidence for the presence of QGS in 2D superconducting system~\cite{SciXing}. Thereafter, QGS was observed in many 2D superconducting systems, such as LaAlO$_3$/SrTiO$_3$ interface~\cite{prbShen}, monolayer NbSe$_2$ films~\cite{NLXing}, ion-gated ZrNCl and MoS$_2$~\cite{NcSaito}. More recently, the existence of QGS has also been confirmed in 2D InO$_x$~\cite{prbLewellyn}, WSi~\cite{CpZhang}, $\beta$-W~\cite{SbHuang}, Li$_x$MoS$_2$~\cite{2dmVerzhbitskiy}, and NbN films~\cite{prbJing, prbWang}. In addition, it has been reported that in four-monolayer crystalline PbTe$_2$ films the QGS  can emerge not only in perpendicular fields but also in parallel magnetic fields~\cite{prlLiu}. According to previous reports~\cite{SciXing, prbShen, NcSaito, NLXing}, quenched disorder is the main origin of the QGS. On the other hand, the theoretical results indicate that the QGS can occur in both 2D and 3D systems~\cite{prlGriffths, prbMotrunich, prlPich, PreVojta, prbKovacs, prlDelMaestro, prbFan}, and the QGS has been observed in 3D magnetic materials~\cite{prldeAndrade, prlCastroNeto, SciSteppke, prlUbaidKassis}. However, there is almost no experimental evidence for the existence of QGS in 3D superconducting system thus far~\cite{note1, note2}. Therefore, it is desirable to explore whether there is QGS in 3D superconductors.

MoN-based films with NaCl structure could be a suitable system to realize the QGS in 3D superconducting films. In 1980s, it was predicted that the superconducting transition temperature $T_{\rm c}$ of MoN with NaCl structure (so called $B$1-MoN) could be as high as 29\,K~\cite{PrbcPickett, sscZhao}. However, the predicted high transition temperature has not yet been achieved in $B$1-MoN probably due to its structural instability (the existence of many Mo or nitrogen vacancies, dislocations and other defects)~\cite{prbIhara, prbInumaru, jacOzsdolay, prbPapaconstantopoulos, prbShi, prbHart}. In addition, it has been found that adding Ti to $B$1-MoN can stabilize its structure~\cite{prbSanjines}. Therefore, to explore whether the $T_{\rm c}$ of Ti stabilized \emph{B}1-MoN is enhanced, we fabricated a series of $B$1-Mo$_{0.8}$Ti$_{0.2}$N$_x$ films with different nitrogen contents $x$. Although the superconducting transition temperature of the MoTiN films were not significantly improved as expected, it is found that 3D MoTiN films possess relatively high $T_{\rm c}$ and low upper critical magnetic field (see in the following text). These features, together with the nature of easily formed structural defects, could make MoTiN films be a suited system for exploring the characteristics of 3D QGS. In the present paper, the low-temperature electrical transport properties of a series of $\sim$80-nm-thick Mo$_{0.8}$Ti$_{0.2}$N$_x$ films with $x$ ranging from $\sim$0.53 to $\sim$1.36 were studied. The QGS is found emerging in the $0.84\lesssim x \lesssim 1.12$ films. We present and discuss the interesting observations in the following subsections.

\section{Experimental method}
Our MoTiN films with thickness $t\sim80$\,nm were grown on (100) MgO single crystal substrates by the reactive magnetron sputtering method. A MoTi alloy target with Mo/Ti ratio of $8:2$ and purity of 99.9\% was selected as the sputtering source. The base pressure of the chamber was less than $1 \times 10^{-4}$\,Pa. During the deposition, the sputtering power was set as 300\,W, and the substrate temperature and sputtering pressure are kept at 500\,$^\circ$C and 0.15\,Pa, respectively. The deposition was carried out in a mixture of argon (99.999\%) and nitrogen (99.999\%) atmosphere. To obtain MoTiN films with different superconducting transition temperatures, the volume ratio of nitrogen to argon was controlled in each deposition round. For the films used in this paper, the volume ratios were set as $1:9$, $1:7$, $3:17$, $7:33$, $1:4$, and $1:3$ (corresponding to nitrogen partial pressures $P_{\rm N_2}\simeq 10.0\%$, 12.5\%, 15.0\%, 17.5\%, 20.0\%, and 25.0\%), respectively.

The thicknesses of the film was controlled by growth rate and deposition time, and further determined by the high-resolution transmission electron microscopy (HRTEM) of the cross section of the film. The crystal structure was determined by X-ray diffraction (XRD), including normal $\theta$-2$\theta$, $\phi$, and $\omega$ scans. The composition of the film and valence of each element were measured using a x-ray photoelectron spectrometer (Thermo Scientific Escalab 250Xi). Before the x-ray photoelectron spectroscopy (XPS) measurement was carried out, the surface of each film was etched by Ar ions and the corrosion depth was $\sim$20\,nm. The microstructure of the films was characterized by transmission electron microscopy (TEM, Tecnai G2 F20 S-Twin). The longitudinal and Hall resistance vs temperature and magnetic field was measured using the standard four-probe method in a physical properties measurement system (PPMS-6000, Quantum Design) equipped with a $^3$He refrigerator. Hall-bar-shaped films (1.0\,mm wide, 10.0\,mm long, and 3.0\,mm distance between the two electrodes) defined by mechanical masks were used in the measurements. For the longitudinal resistance vs magnetic field at a fixed temperature and longitudinal resistance vs temperature at a fixed field measurements, the field was applied in directions being perpendicular and parallel to the film plane, respectively. In the latter case, the field was also perpendicular to the excitation current.

\begin{figure}[htp]
\includegraphics [scale=1.05]{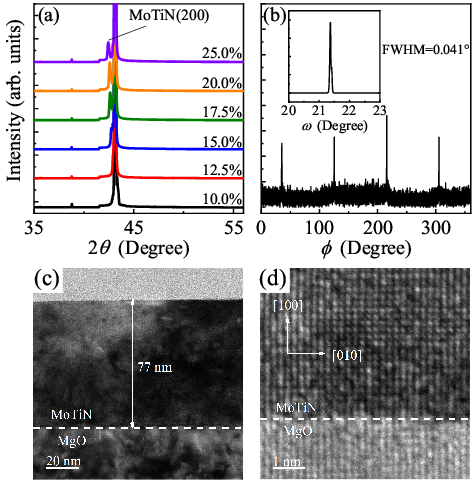}
\caption{\label{figXRD}(a) XRD $\theta$-2$\theta$ scan patterns of MoTiN films deposited at different nitrogen partial pressures. (b) $\phi$-scan spectrum of (220) plane for the film deposited at $P_{\rm N_2} \simeq 17.5\%$. The inset in (b) is the rocking curve of the (200) diffraction peak of the $P_{\rm N_2} \simeq 17.5\%$ film. (c) Cross-sectional HRTEM micrograph of the film deposited at $P_{\rm N_2} \simeq 17.5\%$. (d) The enlarged HRTEM image near the MoTiN/MgO interface for (c).}
\end{figure}

\begin{figure*}[htpb]
\includegraphics[scale=1.45]{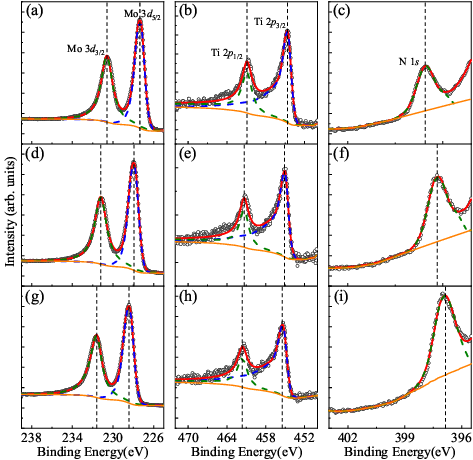}
\caption{\label{figXPS}XPS results of the films deposited at $P_{\rm N_2} \simeq 10.0\%$, 17.5\%, and 25.0\%. (a), (b), and (c) are the spectra of Mo-3$d$, Ti-2$p$, and N-1$s$ orbits for the $P_{\rm N_2} \simeq 10.0\%$ film. (d), (e), and (f) are the spectra of Mo-3$d$, Ti-2$p$, and N-1$s$ orbits for the $P_{\rm N_2} \simeq 17.5\%$ film. (g), (h), and (i) are the spectra of Mo-3$d$, Ti-2$p$, and N-1$s$ orbits for the $P_{\rm N_2} \simeq 25.0\%$ film.}
\end{figure*}

\section{Results and Discussions}\label{SecIII}
In this section, we will first discuss the structures, atomic valences, and compositions of the MoTiN films. Then we will explore the fundamental transport properties of the films. Finally, we will study the characteristics of the quantum phase transition driven by magnetic field. In the following discussions, we use $\perp$ ($\parallel$) as the superscript or subscript of a physical quantity to represent that the quantity is related to the field perpendicular (parallel) to the film plane.
\subsection{Crystal structure and atomic valence}
Figure~\ref{figXRD}(a) shows the XRD $\theta$-$2\theta$ scan patterns of the films deposited at different nitrogen partial pressures. In the spectra, the strong peak centered at $\sim$43.20$^{\circ}$ and the weak peaks at $\sim$38.78$^{\circ}$ are the diffractions of MgO (200) plane. The former corresponds to Cu $K_{\alpha}$ radiation, and the latter is related to Cu $K_\beta$ radiation. For each film, besides the (200) peaks of the MgO substrate,  only the (200) diffraction of the face-centered cubic (fcc) MoTiN can be observed. The values of the lattice constant $a$ can be evaluated using the position of (200) diffractions. For the film deposited at $P_{\rm N_2} \simeq 10.0\%$, the value of $a$ is 4.191\,\AA, which is close to that of $\gamma$-Mo$_2$N~\cite{prbIhara}.  For other films, the values of $a$ increase from 4.228 to $4.272\,{\rm \AA}$  as $P_{\rm N_2}$ increases from 12.5\% to 25.0\%. These values of the lattice constants are comparable with those of $B$1-MoN in previous reports~\cite{jpfLinker, PrbcPickett, sscZhao}. Since the ionic radius of Ti$^{3+}$ is quite close to that of Mo$^{3+}$~\cite{ActaCryst-1976}, it is deduced that MoTiN films with $B$1-structure could be formed under  $12.5\% \lesssim P_{\rm N_2}\lesssim 25.0\%$. Figure~\ref{figXRD}(b) shows the $\phi$-scan profile of (220) plane of the film deposited at $P_{\rm N_2} \simeq 17.5\%$. Four uniformly distributed diffraction peaks can be clearly observed, indicating that the MoTiN film is epitaxially grown on the MgO substrate. The inset of Fig.~\ref{figXRD}(b) shows the rocking curve of the (200) diffraction peak of the $P_{\rm N_2} \simeq 17.5\%$ film. The full width at half maximum (FWHM) of the peak is 0.041$^\circ$, demonstrating that the film has high crystalline quality. For other films, the $\phi$-scan profiles and rocking curves are similar to those of the $P_{\rm N_2} \simeq 17.5\%$ film, and the FWHMs of the peaks in the rocking curves are less than 0.12$^\circ$. The lattice constant and FWHM of the rocking curve for each film are listed in Table~\ref{tab-1}. Figure~\ref{figXRD}(c) shows the cross-sectional HRTEM micrograph of the $P_{\rm N_2}\simeq 17.5\%$ film along the [001] axis. The MoTiN/MgO interface and the surface of the MoTiN film can be clearly identified from the figure. The thickness of the film estimated from the TEM image is $\sim$77\,nm, which is roughly identical to that obtained by the growth rate and deposition time. Figure~\ref{figXRD}(d) shows the enlarged HRTEM image near the MoTiN/MgO interface for the same film. The spacing $d$ between adjacent (100) [or (010)] planes obtained from HRTEM is 0.213\,nm, which is consistent with the result from XRD. Combining with the XRD results, one  further confirms that the $\langle 100\rangle$-oriented MoTiN  films are epitaxially grown on the (100) MgO substrates.

\begin{table*}[htbp]
\caption{
Relevant parameters of the Mo$_{0.8}$Ti$_{0.2}$N$_x$ films. Here $a$ is the lattice constant, FWHM is the full width at half maximum in the rocking curve, $\rho(300\,{\rm K})$ is the resistivity at 300\,K, $T_{\rm c}$ is the superconducting transition temperature, $n(10\,{\rm K})$ is the carrier concentration at 10\,K, $k_{\rm F}\ell$ is the Ioffe-Regel parameter at 10\,K, $B_{\rm c2}(0)$ is the upper critical magnetic field at 0\,K, and $\xi_{\rm GL}$ is the Ginzburg-Landau coherence length. The superscripts $\perp$ and $\parallel$ represent the related quantities obtained for perpendicular and parallel field, respectively.    }\label{tab-1}
\begin{ruledtabular}
\begin{tabular}{ccccccccccccc}
Film    & $P_{\rm N_2}$ &  $a$  & FWHM  & $x$ & $\rho(300\,{\rm K})$ & $T_{\rm c}$  & $n(10\,{\rm K})$ & $k_{\rm F}\ell$ & $B_{\rm c2}^{\perp}(0)$ & $B_{\rm c2}^{\parallel}(0)$ & $\xi_{\rm GL}^{\perp}$ & $\xi_{\rm GL}^{\parallel}$ \\
No.  &  (\%) & (${\rm \AA}$) & (degree) &     & (m$\Omega$\,cm)    & (K)   & ($10^{23}\,{\rm cm}^{-3}$) &   & (T)             & (T) &(nm)  &  (nm)   \\
\hline
1 & 10.0 & 4.191 & 0.048 & 0.53 & 1.245 & 3.30 &  2.53 & 2.78 & 5.21 & 7.38 & 7.96 & 6.68  \\
2 & 12.5 & 4.228 & 0.046 & 0.84 & 1.267 & 3.75 &  2.47 & 2.69 & 6.05 & 8.21 & 7.38 & 6.34\\
3 & 15.0 & 4.232 & 0.115 & 1.05 & 1.284 & 3.13 &  2.41 & 2.55 & 5.23 & 7.27 & 7.94 & 6.73 \\
4 & 17.5 & 4.255 & 0.045 & 1.12 & 1.904 & 2.15 &  2.03 & 1.63 & 3.49 & 5.73 & 9.72 & 7.59\\
5 & 20.0 & 4.257 & 0.043 & 1.20 & 1.942 &      &  1.97 & 1.56 &      &    &      &   \\
6 & 25.0 & 4.272 & 0.049 & 1.36 & 1.942 &      &  1.85 & 0.76 &      &    &      &   \\
\end{tabular}
\end{ruledtabular}
\end{table*}

Figure~\ref{figXPS} shows the XPS results of the film deposited at $P_{\rm N_2} \simeq 10.0\%$, 17.5\%, and 25.0\% as examples. Among them, Fig.~\ref{figXPS}(a), \ref{figXPS}(b), and \ref{figXPS}(c) are the spectra for the $P_{\rm N_2} \simeq 10.0\%$ film, Fig.~\ref{figXPS}(d), \ref{figXPS}(e), and \ref{figXPS}(f) are the spectra for the $P_{\rm N_2} \simeq 17.5\%$ film, and Fig.~\ref{figXPS}(g), \ref{figXPS}(h), and \ref{figXPS}(i) are the spectra for the $P_{\rm N_2} \simeq 25.0\%$ film. For the Mo 3$d$ spectra, the characteristic peaks can be deconvoluted into two peaks, originating from Mo 3$d_{\rm 3/2}$ and Mo 3$d_{\rm 5/2}$ of fcc (MoTi)N$_x$ (Mo$^{\rm \delta+}$, $2<{\rm \delta}<4$), respectively~\cite{assKim, msebWang}. The Ti 2$p$ spectra can also be deconvoluted into two peaks, whose positions are close to those in fcc TiN$_x$~\cite{japDelfino}. For the film deposited at $P_{\rm N_2} \simeq 17.5\%$, the binding energy of the Mo 3$d_{\rm 3/2}$ peak is 231.20\,eV, which is nearly identical to that of $B$1-MoN~\cite{prbInumaru}. As the nitrogen partial pressure is enhanced from 12.5\% to 25.0\%, the binding energies of Mo 3$d_{\rm 3/2}$ peaks gradually increase from 230.85\,eV to 231.60\,eV, which is similar to the results in previous reports~\cite{prbInumaru, msebWang}. The continuous shift of the Mo 3$d_{\rm 3/2}$ peaks to the high binding energy can be attributed to the increase of the charge transfer from Mo to N with the increase of the amount of nitrogen atoms in (MoTi)N$_x$ lattice. For the film deposited at $P_{\rm N_2}\simeq 10.0\%$, the Mo 3$d_{\rm 3/2}$ peak is located at 230.60\,eV, which is decreased about 0.25\,eV compared with that of the $P_{\rm N_2}\simeq 12.5\%$ film. The relative large reduction in binding energy means a great increase of nitrogen vacancies in the $P_{\rm N_2}\simeq 10.0\%$ film. The variation trend of the position of Ti 2$p_{3/2}$ peak is similar to that of the Mo 3$d_{\rm 3/2}$, and the position of Ti 2$p_{3/2}$ peak increases from 454.64\,eV to 455.35\,eV as $P_{\rm N_2}$ is enhanced from 10.0\% to 25.0\%. As for the N $1s$ spectra, the binding energy of the peak decreases with increasing $P_{\rm N_2}$, which is also the result of increasing charge transfer from Mo to N.  From the areas of the XPS peaks, one can estimate the atomic ratios of Mo, Ti, and N in the films. For the films deposited at $P_{\rm N_2} \simeq 10.0\%$, 12.5\%, 15.0\%, 17.5\%, 20.0\%, and 25.0\%, the atomic ratios of Ti to the sum of Mo and Ti are 17.9\%, 19.9\%, 18.3\%, 18.2\%, 17.8\%, and 18.8\%, respectively, i.e., the atomic ratio of Ti to Mo in each film is close to that of the target. In the following discussion, the atomic ratio of Mo to Ti is approximately written as $8:2$ for simplicity. The atomic ratio of the metals (Mo and Ti) to nitrogen, denoted by the subscript $x$ in (MoTi)N$_x$, is also obtained, and listed in Table~\ref{tab-1}. Inspection of Table~\ref{tab-1} indicates that $x$ increases from 0.53 to 1.36 as the nitrogen partial pressure is enhanced from 10.0\% to 25.0\%. The ratio of metal to nitrogen atoms for the film deposited at $P_{\rm N_2} \simeq 10.0\%$ is close to $2:1$, and that for the film deposited at $P_{\rm N_2} \simeq 15.0\%$ is approximately $1:1$. For the molybdenum-nitrogen system, the fcc $\gamma$-Mo$_2$N$_y$ with $0.78\lesssim y \lesssim 1.08$ is known as a stable phase~\cite{prbKanoun}, while $B$1-MoN is believed to be a metastable phase~\cite{jlcm}. The XRD and XPS results discussed above indicate that MoTi-N system can not only form fcc $\gamma$-(Mo$_{0.8}$Ti$_{0.2}$)$_2$N phase but also be stabilized in $B$1-Mo$_{0.8}$Ti$_{0.2}$N$_x$ phase in a wide range of $x$. Our results are consistent with those obtained by Sanjin\'{e}s \emph{et al}~\cite{prbSanjines}.

\begin{figure}[htbp]
\includegraphics [scale=0.8]{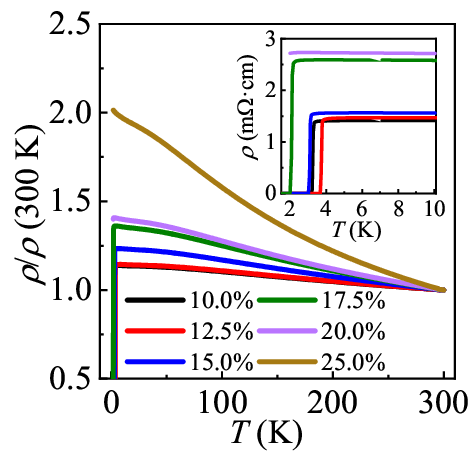}
\caption{\label{figRT} Normalized resistivity as a function of temperature for the MoTiN films deposited at different nitrogen partial pressures. Inset: Resistivity vs temperature at low temperature regime for films deposited at $P_{\rm N_2} \lesssim 20.0\%$.}
\end{figure}

\subsection{Fundamental transport properties}
Figure~\ref{figRT} shows the normalized resistivity $\rho / \rho(300\,{\rm K})$ as a function of temperature $T$ for the films deposited at different nitrogen partial pressures. The resistivities of all the films slightly increase with decreasing temperature above liquid helium temperature, i.e., the temperature coefficient of the resistivity $(1/\rho)(\mathrm{d}\rho/\mathrm{d}T)$ for each film is negative at high temperature regime. The films deposited at $P_{\rm N_2} \lesssim 17.5\%$ enter into superconducting state below the superconducting transition temperature $T_{\rm c}$, where $T_{\rm c}$ is designated as the temperature at which the resistance drops to 90\% of that in normal state [$\rho(10\,{\rm K})$, see the inset of Fig.~\ref{figRT}]. For the $B$1-Mo$_{0.8}$Ti$_{0.2}$N$_x$ films, the superconducting transition temperature increases with decreasing nitrogen partial pressure, while the $T_c$ of the $P_{\rm N_2} \simeq 10.0\%$  [$\gamma$-(Mo$_{0.8}$Ti$_{0.2}$)$_2$N$_{1.06}$] film is less than that of the $P_{\rm N_2} \simeq 12.5\%$ [$B$1-Mo$_{0.8}$Ti$_{0.2}$N$_{0.84}$] film. Using free-electron model, one can obtain the product of the Fermi wave number and mean free path of electrons $k_{\rm F} \ell$ via $k_{\rm F} \ell = (\hbar/e^2)(3\pi^2 )^{1/3}n^{-2/3}/\rho$, where $\hbar$ is the Planck's constant divided by $2\pi$, $e$ is the electronic charge, and $n$ is the carrier concentration. The values of $k_{\rm F} \ell$ for the films at 10\,K are summarized in Table~\ref{tab-1}. Inspection of Table~\ref{tab-1} indicates that the $k_{\rm F} \ell$ is 0.76 for the $P_{\rm N_2} \simeq 25.0\%$ film, and varies between 1.56 and 2.78 for other films. According to the Ioffe-Regel criterion~\cite{pgIoffe}, the $P_{\rm N_2} \simeq 25.0\%$ film just lies in the metal-insulator transition region, while the other films are on the metal side of the metal-insulator transition and in a weakly localized metallic state.

\begin{figure}[htbp]
\includegraphics [scale=1.05] {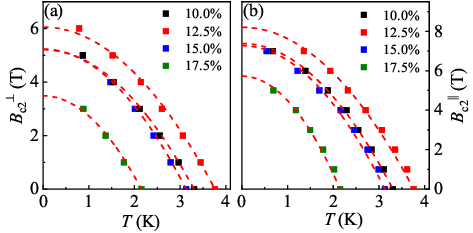}
\caption{\label{fig-CohLen}(a) The perpendicular upper critical magnetic field $B_{\rm c2}^{\perp}$ vs temperature $T$ for the films deposited at $P_{\rm N_2} \lesssim 17.5\%$. (b) The parallel upper critical magnetic field $B_{\rm c2}^{\parallel}$ vs temperature $T$ for the films deposited at $P_{\rm N_2} \lesssim 17.5\%$.  The dashed curves are the least-squares fits to $B_{\rm c2}^{i}(T)=B_{\rm c2}^{i}(0)[1-(T/T_{\rm c})^2]$ with $i=\perp$ in (a) and $i=\parallel$ in (b).}
\end{figure}

Figure~\ref{fig-CohLen}(a) and \ref{fig-CohLen}(b) show the temperature dependence of the perpendicular and parallel upper critical fields for the $P_{\rm N_2} \lesssim 17.5\%$ films.  The experimental $B_{\rm c2}^{\perp}$-$T$ and $B_{\rm c2}^{\parallel}$-$T$ data are  fitted to $B_{\rm c2}^{i}(T)=B_{\rm c2}^{i}(0)[1-(T/T_{\rm c})^2]$~\cite{Tinkha-book}, where $i=\perp$ and $\parallel$ stand for the perpendicular-field and parallel-field situations, respectively,  and $B_{\rm c2}^{i}(0)$ is the upper critical magnetic field at 0\,K.  The fitted results are shown in Fig.~\ref{fig-CohLen} by the dashed curves. The values of the adjusting parameters $B_{\rm c2}^{\perp}(0)$ and $B_{\rm c2}^{\parallel}(0)$ are obtained and listed in Table~\ref{tab-1}. Inspection of  Table~\ref{tab-1} indicates that $B_{\rm c2}^{\parallel}(0)$ is only slightly greater than $B_{\rm c2}^{\perp}(0)$ for each film, suggesting the films are 3D with respect to superconductivity. Once the values of $B_{\rm c2}^{\perp}(0)$ and $B_{\rm c2}^{\parallel}(0)$ are obtained, one can deduce the Ginzburg-Landau (GL) coherence lengths $\xi_{\rm GL}^{\perp}$ and $\xi_{\rm GL}^{\parallel}$ for each film via $\xi^{i}_{\rm GL}=[\Phi_{\rm 0}/(2\pi B_{\rm c2}^{i}(0))]^{1/2}$ ~\cite{Tinkha-book}, where $\Phi_0 = h/2e$ is the flux quantum. The GL coherence lengths $\xi^{\perp}_{\rm GL}$ and $\xi^{\parallel}_{\rm GL}$ for each superconducting film are also listed in Table~\ref{tab-1}. Clearly, the coherence length perpendicular to the film plane $\xi_{\rm GL}^{\perp}$ lies between $\sim$7 and $\sim$10\,nm, while the coherence length parallel to the film plane $\xi_{\rm GL}^{\parallel}$ varies  between $\sim$6 and $\sim$8\,nm. For each film, $\xi_{\rm GL}^{\perp}$ is comparable with $\xi_{\rm GL}^{\parallel}$ and much less than the thickness of the film, which further confirms the 3D superconductive characteristics of the MoTiN films.

\begin{figure}[htbp]
\includegraphics [scale=1.05] {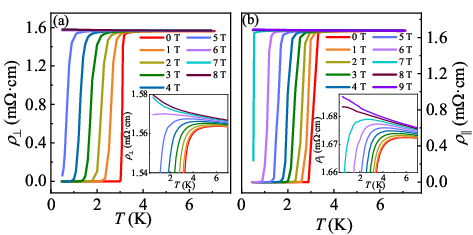}
\caption{\label{figBRT}(a) The resistivity as a function of temperature in different fields (perpendicular to the film plane) for the film deposited at $P_{\rm N_2} \simeq 15.0\%$. (b) The resistivity as a function of temperature in different fields (parallel to the film plane) for the film deposited at $P_{\rm N_2} \simeq 15.0\%$. The insets in (a) and (b) are the enlarged views of the corresponding films near the superconducting transition region. }
\end{figure}

\subsection{Quantum phase transition}
In this subsection, we will focus on the quantum phase transitions in these 3D superconducting films deposited at $P_{\rm N_2} \lesssim 17.5\%$. We first explore the temperature dependent behavior of the resistivity evolution with external field. Considering the results for the $P_{\rm N_2}\simeq17.5\%$, 15.0\%, 12.5\%, and 10.0\% films are similar, we only present and discuss the results for a representative film deposited at $P_{\rm N_2}\simeq15.0\%$. Figure~\ref{figBRT}(a) shows the resistivity variation with temperature from 7.00 down to 0.50\,K under different perpendicular fields for the $P_{\rm N_2}\simeq15.0\%$ film. The inset of this figure is the close view of the superconducting transition region. When the field is less than $\sim$6\,T, the film transforms from normal state to superconducting state upon cooling and the superconducting transition temperature decreases with increasing field. As the magnetic field is increased to 7\,T, the resistivity slightly increases with decreasing temperature over the whole measured temperature range. Considering the $k_{\rm F}\ell$ of the film is 2.55 at 10\,K, one can deduce that the film transforms into a weakly localized metallic state under 7\,T. Inspection of the inset of Fig.~\ref{figBRT}(a) also indicates that the critical filed for superconductor-metal transition (SMT) is in the vicinity of $\sim$6.0\,T. Figure~\ref{figBRT}(b) shows the resistivity vs temperature at different parallel fields for the same film. Similar to the situation at perpendicular field, the parallel field also drives the film from a superconducting state to a weakly localized metallic state. The parallel critical filed for SMT lies in the vicinity of $\sim$8.0\,T, being slightly larger than the perpendicular critical filed.

For field induced SMT or superconductor-insulator transition (SIT), the dirty-boson model predicts that the field dependence of resistance curves at different temperatures all cross at a single point with $B=B_{\rm c}$ ($B_{\rm c}$ is the critical field)~\cite{prlYazdani, prbSteiner, pnasBreznay}. In addition, near the SMT (SIT) the resistance at different fields and temperatures obeys a power-law scaling form~\cite{prlFisher, prbFisher}
\begin{equation}\label{Eq-scaling}
 R(B, T)=R_{\rm c} f(\delta T^{-1/z\nu}),
\end{equation}
where $R_{\rm c}$ is the critical resistance,$f(x)$ is the scaling function with $f(0)=1$, $\delta=|B-B_{\rm c}|$ is the distance from the critical field $B_{\rm c}$, $\nu$ is the correlation length exponent, $z$ is the dynamical critical exponent. Experimentally, the power-law scaling in Eq.~(\ref{Eq-scaling}) has been observed in a range of 2D superconductors near the SIT or SMT. The values of $z\nu$ were generally found to be $\approx$0.65~\cite{prbMarrache-Kikuchi}, $\approx$1.33~\cite{prlYazdani, prlMason}, and $\approx$2.33~\cite{prbSteiner, pnasBreznay}, which correspond to the universality class of $(2 + 1)$D\,$XY$ model~\cite{rmbSondhi}, classic percolation model~\cite{prbSteiner}, and quantum percolation model~\cite{prbSteiner}, respectively. As mentioned in Sec.~\ref{secI}, in some 2D superconducting films, the magnetoresistance isotherms do not intersect at a single point or narrow region, but rather in a wide region, indicating the presence of QGS accompanying the SMT.

\begin{figure}[htbp]
\includegraphics[scale=1.05]{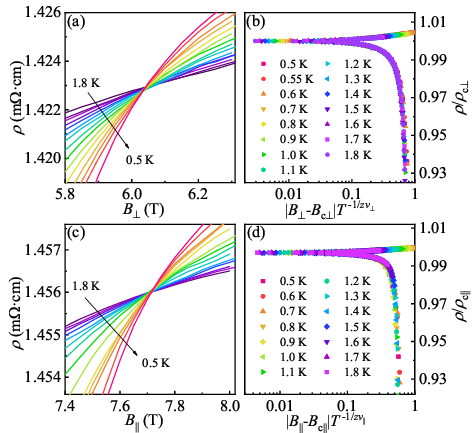}
\caption{\label{FigMo2N}(a) Resistivity versus perpendicular field at different temperatures ranging from 0.50 to 1.80\,K for the $P_{\rm N_2}\simeq10.0\%$ film. (b) Normalized resistivity $\rho/\rho_{\rm c\perp}$ as a function of the
scaling variable $|B_{\perp}-B_{\rm c\perp}|T^{-1/z\nu_{\perp}}$ measured at perpendicular field. (c) Resistivity versus parallel field at different temperatures ranging from 0.50 to 1.80\,K for the $P_{\rm N_2}\simeq10.0\%$ film. (d)  Normalized resistivity $\rho/\rho_{\rm c\parallel}$ as a function of the scaling variable $|B_{\parallel}-B_{\rm c\parallel}|T^{-1/z\nu_{\parallel}}$ measured at parallel field.}
\end{figure}

\begin{table*}[htbp]
\caption{
Relevant parameters of the Mo$_{0.8}$Ti$_{0.2}$N$_x$ films deposited at $P_{\rm N_2}\simeq12.5\%$, 15.0\%, 17.5\%. Here $C$ and $B_{\rm c}^\ast$ are the parameters in  Eq.~(\ref{Eq-Infinite-Randomness-1}); $\tilde{B}_{\rm c}^{\ast}$, $\tilde{T}_0$, $u$, and $y$ are the characteristic field, characteristic temperature and parameters in Eq.~(\ref{Eq-Infinite-Randomness-2}). The subscripts $\perp$ and $\parallel$ represent the related quantities obtained for perpendicular and parallel field, respectively. }\label{tab-2}
\begin{ruledtabular}
\begin{tabular}{cccccccccccccc}
Film    & $P_{\rm N_2}$ &  $C_{\perp}$ & $B_{\rm c \perp}^\ast$ & $\tilde{B}_{\rm c \perp}^{\ast}$ & $\tilde{T}_{0 \perp}$ & $u_{\perp}$ & $y_{\perp}$ &  $C_{\parallel}$ & $B_{\rm c \parallel}^\ast$ & $\tilde{B}_{\rm c \parallel}^{\ast}$ & $\tilde{T}_{0 \parallel}$ & $u_{\parallel}$ & $y_{\parallel}$\\
No.     &  (\%)         &      &  (T)             & (T)                     & (K)          &     &   &      &  (T)             & (T)                     & (K)          &     &    \\
\hline
2 & 12.5 &   0.318 & 7.341 & 7.469 & 5.24 & 0.084 & 0.02&   0.379 & 9.046 & 9.279 & 4.725 & 0.096 & 0.03\\
3 & 15.0 &   0.284 & 6.170 & 6.155 & 3.446 & 0.0256 & 0.008&   0.434 & 7.962 & 8.075 & 3.314 & 0.063 & 0.012\\
4 & 17.5 &   0.291 & 4.595 & 4.682 & 2.703 & 0.064 & 0.16&   0.482 & 6.287 & 6.499 & 3.606 & 0.205 & 0.24\\
\end{tabular}
\end{ruledtabular}
\end{table*}

To explore whether the QGS could occur in 3D superconductors, we measured the magnetoresistance isotherms of the MoTiN films deposited at $P_{\rm N_2} \lesssim 17.5\%$. Figure~\ref{FigMo2N}(a) shows the low-temperature magnetoresistance isotherms of the (Mo$_{0.8}$Ti$_{0.2}$)$_{2}$N$_{1.06}$ film (i.e., the film deposited at $P_{\rm N_2}\simeq 10.0\%$) measured at fields perpendicular to the film plane. Clearly, the $\rho$-$B_{\perp}$ curves measured at different temperatures almost cross at a single point. The field and resistivity corresponding to the crossing point are $B_{\rm c \perp} \approx 6.04$\,T and $\rho_{\rm c \perp} \approx 1.423$\,m$\Omega$\,cm. Using the values of the $\rho_{\rm c \perp }$ and $B_{\rm c\perp}$, we compare the $\rho(B_{\perp},T)/\rho_{\rm c \perp}$ data with Eq.~(\ref{Eq-scaling}). As shown in Fig.~\ref{FigMo2N}(b), the $\rho/\rho_{\rm c \perp}$ vs $|B_{\perp}-B_{\rm c\perp}|T^{-1/z\nu_{\perp}}$ collapse onto a single curve (two branches) as $z\nu_{\perp}$ is taken the appropriate value. The optimized value of $z\nu_{\perp}$ for the best collapse is $z\nu_{\perp}=0.84 \pm 0.05$ using a numerical minimization procedure. For the parallel field case, the low-temperature magnetoresistance isotherms and related scaling curve are shown in Fig.~\ref{FigMo2N}(c) and \ref{FigMo2N}(d), respectively. Clearly, the results are similar to those measured at the perpendicular field: the low-temperature magnetoresistance isotherms cross at a single point and the $\rho(B_{\parallel}, T)/\rho_{\rm c \parallel}$ data obey the power-law scaling form of Eq.~(\ref{Eq-scaling}). Here the optimized scaling parameter $z\nu_{\parallel}$ is $z\nu_{\parallel}=0.74 \pm 0.02$, and the critical resistivity and field are $\rho_{\rm c\parallel} \approx 1.456$\,m$\Omega$\,cm and $B_{\rm c\parallel} \approx 7.72$\,T, respectively. It should be noted that value of $B_{\rm c\parallel}$ is only 1.28 times as large as that of $B_{\rm c\perp}$, which also supports that the (Mo$_{0.8}$Ti$_{0.2}$)$_{2}$N$_{1.06}$ film is 3D with respect to superconductivity. The features of the magnetoresistance isotherms strongly suggest that the Cooper pairs in the 3D (Mo$_{0.8}$Ti$_{0.2}$)$_{2}$N$_{1.06}$ film can also be described by a two-dimensional boson system and the SMT in the film is caused by quantum fluctuations.  In fact, it is found that the $R(B,T)$ data of the 50-nm-thick Nb$_{0.15}$Si$_{0.85}$ superconducting films~\cite{PRB-73-094521} and 3D BaPb$_{1-x}$Bi$_x$O$_3$ ($0.24 \leqslant x \leqslant 0.29$) superconductors~\cite{prbGiraldo-Gallo} in the vicinity of the SIT can also be described by Eq.~(\ref{Eq-scaling}).

\begin{figure}
\includegraphics{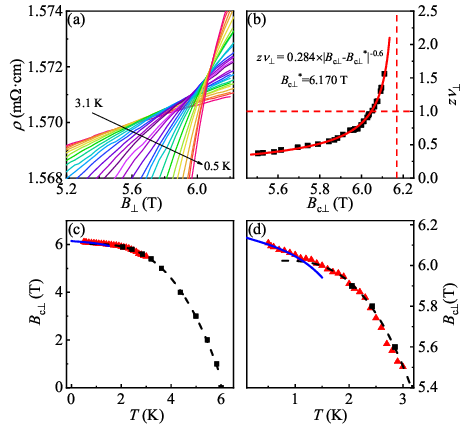}
\caption{\label{figQGS}(a) Resistivity vs perpendicular field at different temperatures ranging from 0.50 to 3.10\,K for the film deposited at $P_{\rm N_2}\simeq15.0\%$. (b) $B_{\rm c \perp}$ dependence of critical exponent $z\nu_\perp$. The dashed vertical and horizontal lines represent the trajectories of $B_{\rm c\perp}= B_{\rm c \perp}^\ast$ and $z\nu_{\perp}=1$, respectively. (c) Temperature dependence of perpendicular critical field (solid triangles) and $T^{\rm on}_{\rm c}$ at different fields (solid squares). The solid curve is the least-square fit to Eq.~(\ref{Eq-Bc-T}) and the dashed curve is only the guide to eyes. (d) The close view of the low temperature and high field region of (c).}
\end{figure}

For the films deposited at $P_{\rm N_2} \simeq 12.5\%$, 15.0\%, 17.5\%, the features of the low-temperature magnetoresistance isotherms, which will be presented below, are quite different from those of the (Mo$_{0.8}$Ti$_{0.2}$)$_{2}$N$_{1.06}$ film. The results for the three films deposited at $P_{\rm N_2} \simeq 12.5\%$, 15.0\%, 17.5\% are similar, hence, we only give and discuss the results taken from the $P_{\rm N_2}\simeq15.0\%$ film. We first explore what happens when the field is perpendicular to the film plane. Figure~\ref{figQGS}(a) shows $\rho$ as a function of $B_{\perp}$ at different temperatures from 0.50 to 3.10\,K. Clearly, the magnetoresistance isotherms at low temperatures do not cross at a single point or a narrow region, but cross at many points located in a relatively large transition area. These crossing points form a continuous boundary line of SMT. This phenomenon is very similar to that in 2D superconductors with QGS~\cite{SciXing, prbShen, NcSaito, NLXing}, and treated as a signature of QGS in 2D superconductors. We analyze the magnetoresistance isotherms in a fashion similar to those in 2D superconducting films with QGS. Assuming that $\rho$-$B_{\perp}$ curves at three adjacent temperatures cross at one point, we compare the $\rho$-$B_{\perp}$ data measured at three adjacent temperatures with the transformed Eq.~(\ref{Eq-scaling}), $\rho(B, t)=\rho_{\rm c} f(\delta t)$ with $t$ being $t=(T/T_0)^{-1/z\nu}$ and $T_0$ being the lowest temperature in each group of $\rho$-$B_{\perp}$ data. Selecting the proper value of $z\nu_{\perp}$, the $\rho$-$B_{\perp}$ data at the three adjacent temperatures will be scaled onto two branches. Thus, the critical exponent $z\nu_{\perp}$ as a function of $B_{\rm c \perp}$, shown in Fig.~\ref{figQGS}(b), was obtained. In the high temperature regime ($T\gtrsim1.20$\,K), $z\nu_{\perp}$ increases slowly with increasing $B_{\rm c \perp}$, while in the low temperature regime ($T\lesssim1.20$\,K), $z\nu_{\perp}$ increases rapidly and tends to diverge as $B_{\rm c \perp}\rightarrow B_{\rm c \perp}^\ast$, where $B_{\rm c \perp}^\ast$ is the characteristic field. The $z\nu_{\perp}$ vs $B_{\rm c \perp}$ data are compared with the equation~\cite{PreVojta}
\begin{equation}\label{Eq-Bc}
 z\nu=C|B_{\rm c}-B_{\rm c}^\ast|^{-\upsilon\psi},
\end{equation}
where $C$ is a constant, the correlation length exponent $\upsilon=1.2$, and the tunneling exponent $\psi=0.5$~\cite{prbKovacs, prbMotrunich}. As shown by the solid curve in Fig.~\ref{figQGS}(b), the $z\nu_{\perp}$ vs $B_{\rm c \perp}$ data of the film satisfy the activated scaling described by Eq.~(\ref{Eq-Bc}). For the $P_{\rm N_2}\simeq15.0\%$ film, the fitted value of $B_{\rm c \perp}^\ast$ is 6.170\,T. The values of $B_{\rm c \perp}^\ast$ for the other two films are listed in Table~\ref{tab-2}. We note in passing that the theoretical curve of Eq.~(\ref{Eq-Bc}) would deviate from the $z\nu_{\perp}$ vs $B_{\rm c\perp}$ data for each film if we set $\upsilon\psi\simeq0.33$, which is different from that in Ref.~\cite{note2}.

Figure~\ref{figQGS}(c) shows the temperature dependence of perpendicular critical magnetic field for the $P_{\rm N_2}\simeq 15.0\%$ film. Here the solid triangles represent the critical field $B_{\rm c \perp}(T)$ at each crossing point of two adjacent magnetoresistance isotherms, and the solid squares are the superconducting onset temperature $T_{\rm c}^{\rm on}(B_{\perp})$ obtained from the $\rho$-$T$ curves (the temperature for $\mathrm{d}\rho/\mathrm{d}T=0$ near the superconducting transition). It should be noted that the $T_{\rm c}^{\rm on}(B_{\perp})$ data and the $B_{\rm c \perp}(T)$ data almost follow the same trajectory above $\sim$1.10\,K [see the dashed line in Fig.~\ref{figQGS}(c)], which is also similar to that for 2D superconductors with QGS~\cite{SciXing,prbShen,NLXing,NcSaito}. The value of $B_{\rm c \perp}(T)$ gradually increases with decreasing temperature from 3.10 to $\sim$1.10\,K, and then rapidly increases with further decreasing temperature [see Fig.~\ref{figQGS}(d)], i.e., the $B_{\rm c \perp}$ vs $T$ data deviates from its original trajectory below $\sim$1.10\,K. In 2D superconductors with QGS, the temperature dependence of $B_{\rm c}$ in lower temperature regime also deviates from its higher temperature trajectory, which is considered as another signature for the occurrence of QGS~\cite{SciXing}. The similarities of the low-temperature magnetoresistance isotherms and the temperature dependent behavior of $B_{\rm c}$ between the 3D Mo$_{0.8}$Ti$_{0.2}$N$_{x}$ superconducting films and 2D superconductors with QGS, together with the fact that the $z\nu_{\perp}$ vs $B_{\rm c \perp}$ of the Mo$_{0.8}$Ti$_{0.2}$N$_{x}$ films obey the activated scaling law, strongly suggest the occurrence of QGS in the 3D Mo$_{0.8}$Ti$_{0.2}$N$_{x}$ films.

\begin{figure}
\includegraphics{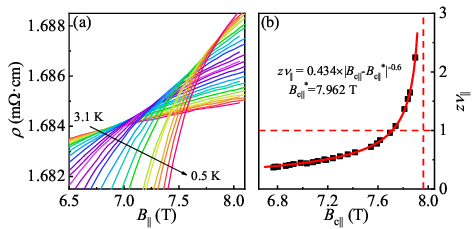}
\caption{\label{figQGSparallel}(a) The resistivity versus the parallel field at different temperatures from 0.50 to 3.10\,K for the $P_{\rm N_2}\simeq15.0\%$ film. (b) $B_{\rm c \parallel}$ dependence of critical exponent $z\nu_\parallel$ for the  $P_{\rm N_2}\simeq15.0\%$ film. The dashed vertical and horizontal lines stand for the trajectories of $B_{\rm c\parallel} =B_{\rm c \parallel}^\ast$ and $z\nu_{\parallel}=1$, respectively.}
\end{figure}

Figure~\ref{figQGSparallel}(a) shows the magnetoresistance isotherms measured at fields being parallel to the film plane for the $P_{\rm N_2}\simeq15.0\%$ film. Similar to those in the perpendicular fields, the low-temperature $\rho$-$B_\parallel$ curves at different temperatures do not cross at a single point or a narrow region, but at a wide range in which the field spans from $\sim6.7$ to $\sim7.9$\,T as the temperature decreases from 3.10 to 0.50\,K. The features of the magnetoresistance isotherms strongly suggest that the QGS may occur in the film as subjected to a parallel field. To verify the speculation, we analyze the magnetoresistance isotherms measured at parallel field, just like analyzing that for the perpendicular field mentioned above. Then, $z\nu_{\parallel}$ as a function of the critical field $B_{\rm c\parallel}$ is obtained, and shown in Fig.~\ref{figQGSparallel}(b). From Fig.~\ref{figQGSparallel}(b), one can see that the variation of $z\nu_{\parallel}$ with $B_{\rm c\parallel}$ is quite similar to that for $z\nu_{\perp}$ with $B_{\rm c\perp}$, i.e., the $z\nu_{\parallel}$ increases with increasing $B_{\rm c\parallel}$ (or decreasing temperature), and tends to diverge as $T\rightarrow 0$\,K. The solid curve in Fig.~\ref{figQGSparallel}(b) is the theoretical prediction of Eq.~(\ref{Eq-Bc}), where the optimized parameters $C_{\parallel}$ and $B_{\rm c\parallel}^{\ast}$ are 0.434 and 7.962\,T, respectively. Inspection of Fig.~\ref{figQGSparallel}(b) indicates that the $z\nu_{\parallel}$ vs $B_{\rm c\parallel}$ data can also be well described by Eq.~(\ref{Eq-Bc}), just like the situation in perpendicular fields mentioned above. Thus, our results suggest that both the perpendicular and parallel fields can drive the $12.5 \% \lesssim P_{\rm N_2}\lesssim 17.5\%$ films into a QGS state before entering into a weakly localized metallic state. Considering the values of the characteristic fields $B_{\rm c\perp}^{\ast}$ and $B_{\rm c\parallel}^{\ast}$ are comparable for each film, one can argue that the QGS is observed in 3D MoTiN superconductors.

\begin{figure}[htbp]
\includegraphics [scale=1.05]{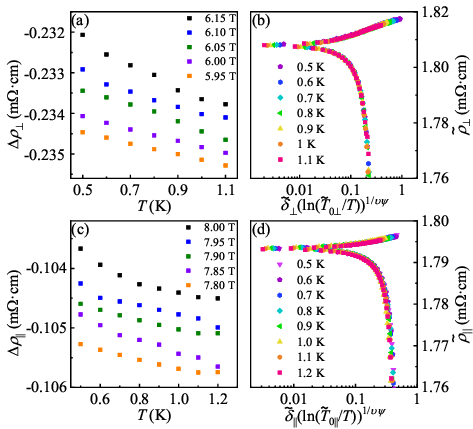}
\caption{\label{figQGS2}The activated scaling analysis with the irrelevant correction for the $P_{\rm N_2}\simeq15.0\%$ film. (a) Temperature dependence of the irrelevant correction at different perpendicular fields. (b) The effective resistivity $\tilde{\rho}$  vs the scaling variable $\tilde{\delta}_{\perp}[\ln(\tilde{T}_{0\perp}/T)]^{1/\upsilon \psi}$. (c) Temperature dependence of the irrelevant correction at different parallel fields. (d) The effective resistivity $\tilde{\rho}$ vs the scaling variable $\tilde{\delta}_{\parallel}[\ln(\tilde{T}_{0\parallel}/T)]^{1/\upsilon \psi}$. Here $\upsilon\psi$ is taken as 0.6 for both the perpendicular and parallel fields.}
\end{figure}

On the other hand, based on the random transverse field Ising model, Maestro \emph{et al}.~\cite{prlDelMaestro} have proposed an activated scaling to describe the conductivity of nanowires near SMT. Recently, it has been demonstrated that the activated scaling can describe the variation of the resistance with the field and temperature near the SMT for the 2D superconductors with QGS~\cite{prbLewellyn, jpdCui}. According to Maestro \emph{et al}.~\cite{prlDelMaestro} and Lewellyn \emph{et al}.~\cite{prbLewellyn}, the activated dynamics scaling can be written as
\begin{equation}\label{Eq-Infinite-Randomness-1}
  R\left( \tilde{\delta}, \ln\frac{\tilde{T}_0}{T} \right) = \Phi \left[ \tilde{\delta}\left(\ln\frac{\tilde{T}_0}{T}\right)^{1/\upsilon\psi}\right],
\end{equation}
where $\tilde{\delta}=|B-\tilde{B}_{\rm c}^{\ast}|/\tilde{B}_{\rm c}^{\ast}$ is the relative distance from the critical field,  $\tilde{B}_{\rm c}^{\ast}$ is the critical field for $T\rightarrow 0$~\cite{note3}, $\upsilon$ is the correlation length exponent, $\psi$ is the tunneling exponent, and $\tilde{T}_0$ is a microscopic temperature scale associated with the quantum phase transition. In Eq.~(\ref{Eq-Infinite-Randomness-1}), the influence of temperature was not considered. At nonzero temperature, an irrelevant correction to the scaling should be included and the scaling relation can be approximately expressed as~\cite{prbLewellyn, jpdCui}
\begin{eqnarray}\label{Eq-Infinite-Randomness-2}
R\left( \tilde{\delta}, \ln\frac{\tilde{T}_0}{T}, u \right) & \approx & \Phi \left[ \tilde{\delta}\left(\ln\frac{\tilde{T}_0}{T}\right)^{\frac{1}{\upsilon\psi}}\right] \nonumber\\
  &+ & u \left(\ln\frac{\tilde{T}_0}{T}\right)^{-y} \Phi_u \left[ \tilde{\delta}\left(\ln\frac{\tilde{T}_0}{T}\right)^{\frac{1}{\upsilon\psi}}\right], \\ \nonumber
\end{eqnarray}
where $u$ is the leading irrelevant scaling variable, and $y>0$ is the associated irrelevant exponent. Expanding Eq.~(\ref{Eq-Infinite-Randomness-2}) around $\tilde{\delta}=0$, one can obtain
\begin{equation}\label{Eq-Bc-T}
 B_{\rm c}(T)\approx \tilde{B}_{\rm c}^{\ast}\left[1-u \left(\ln \frac{\tilde{T}_0}{T}\right)^{-\frac{1}{\upsilon\psi} -y}\right].
\end{equation}
Thus, defining the effective resistance $\tilde{R}$ as $\tilde{R}=\Phi(\tilde{\delta}[\ln \tilde{T}_0/T]^{1/\upsilon\psi})$, one can determine whether the activated scaling is satisfied by checking the $\tilde{R}$ vs $\tilde{\delta}[\ln \tilde{T}_0/T]^{1/\upsilon\psi}$ curve.
Experimentally, it has been demonstrated that the low-temperature $\tilde{R}$ data at different temperatures and fields in the vicinity of $B_{\rm c}$ of the 2D superconductors with QGS follow the prediction of Eq.~(\ref{Eq-Infinite-Randomness-2}) and collapse onto a single curve (two different branches), which in turn is another hallmark of the emergence of QGS.

To check whether the low-temperature magnetoresistance isotherms of the 3D superconducting MoTiN films deposited at $12.5\% \lesssim P_{\rm N_2} \lesssim 17.5\%$ obey the activated scaling mentioned above, we compare the $B_{{\rm c}i}$ vs $T$  and low-temperature $\rho(T,B_{i})$ data with Eq.~(\ref{Eq-Bc-T}) and Eq.~(\ref{Eq-Infinite-Randomness-2}), respectively, where $i=\perp$ and $\parallel$. Firstly, the theoretical predictions of Eq.~(\ref{Eq-Bc-T}) are fitted to the $B_{{\rm c}i}$ vs $T$ data of the films by taking $\upsilon\psi=0.6$ [for the $P_{\rm N_2}\simeq15.0\%$ film in the perpendicular field, see the solid curve in Fig.~\ref{figQGS}(d) as an example]. Then the fitting parameters $B_{{\rm c}i}^\ast$, $\tilde{T}_{0i}$, $u_{i}$, and $y_{i}$ can be determined (see Table~\ref{tab-2}). Secondly, fixing $\tilde{\delta}_i[\ln \tilde{T}_{0i}/T]^{1/\upsilon\psi}$ at a certain value, we obtain the value of $u_{i} \Phi_u$ in Eq.~(\ref{Eq-Infinite-Randomness-2}) via extracting the slope of $\rho$ vs $(\ln \tilde{T}_{0i}/T)^{-y_i}$ of the films. Thus, the irrelevant correction [second term on the right hand side of Eq.~(\ref{Eq-Infinite-Randomness-2})] as a function of temperature and the effective resistivity [the first term on the right hand side of Eq.~(\ref{Eq-Infinite-Randomness-2})] as a function of $\tilde{\delta_i}[\ln \tilde{T}_{0i}/T]^{1/\upsilon\psi}$ for each film are obtained, respectively. As an example, we present the temperature dependence of the irrelevant correction $\Delta\rho_{\perp}$ [$\Delta\rho_{\parallel}$] and the effective resistivity $\tilde{\rho}_{\perp}$ [$\tilde{\rho}_{\parallel}$] variation with  the scaling parameter $\tilde{\delta}_{\perp}(\ln \tilde{T}_{0\perp}/T)^{1/\upsilon\psi}$ [$\tilde{\delta}_{\parallel}(\ln \tilde{T}_{0\parallel}/T)^{1/\upsilon\psi}$]for the $P_{\rm N_2} \simeq 15.0\%$ film in the temperature range of 0.50 to 1.10\,K [1.20\,K] in Fig.~\ref{figQGS2}(a) [Fig.~\ref{figQGS2}(c)] and \ref{figQGS2}(b) [\ref{figQGS2}(d)], respectively. For this film, $\Delta\rho_{\perp}$ ($\Delta\rho_{\parallel}$) is below 0.235\,m$\Omega$\,cm (0.106\,m$\Omega$\,cm), which is less than 15\% (7\%) of the resistivity at the field and temperature regimes. In addition, Both $\Delta\rho_{\perp}$ and $\Delta\rho_{\parallel}$ decrease with decreasing temperature, being consistent with the theoretical predication. From Fig.~\ref{figQGS2}(b) and \ref{figQGS2}(d) , one can see that the $\tilde{\rho}_{i}$ vs $\tilde{\delta}_{i}[\ln \tilde{T}_{0i}/T]^{1/\upsilon\psi}$ ($i=\perp$ and $\parallel$) data collapse onto a single curve (two branches), satisfying the activated scaling described by Eq.~(\ref{Eq-Infinite-Randomness-2}). Inspection of Table~\ref{tab-2} indicates the difference between $B_{{\rm c}i}^{\ast}$ [obtained from Eq.~(\ref{Eq-Bc})] and  $\tilde{B}_{{\rm c}i}^{\ast}$ [obtained from Eq.~(\ref{Eq-Bc-T})] is less than 3.4\% for each film. The results mentioned here confirm the emergence of QGS in these $12.5\% \lesssim P_{\rm N_2} \lesssim 17.5\%$ films.

Theoretically, the QGS in 3D system is stable and originates from the quenched disorder~\cite{prbMotrunich, PreVojta, prlDelMaestro, prbFan}, which is just like that in 2D system. For the \emph{B}1-type MoN film, it is reported that the fcc structure can be also stable when a lot of vacancies, including both molybdenum and nitrogen vacancies, exist in the fcc conventional cell~\cite{jacOzsdolay, jpcXiao}. In the Mo$_{0.8}$Ti$_{0.2}$N$_{x}$ films, $x$ only represents the average atomic ratio of nitrogen to the sum of Mo and Ti and the  number of transition metal or nitrogen atoms in each fcc conventional cell is generally less than 4 even if $x\simeq 1$, i.e., there are a lot of Mo (or Ti) and nitrogen vacancies in the \emph{B}1-Mo$_{0.8}$Ti$_{0.2}$N$_{x}$ films. These vacancies are the main source of the quenched disorder. At low temperature and high field (in the vicinity of the critical point of the quantum phase transition), the quenched disorder drives the films into an inhomogeneous superconducting state, in which the superconducting islands or droplets (rare regions) coexist with the disordered bulk of the system. The dynamics of the rare regions is very slow and follows the activated scaling rather than the power-law dynamical scaling. Thus, the sharp transition of the SMT is smeared and QGS then emerges in these $12.5\%\lesssim P_{\rm N_2}\lesssim 17.5\%$ Mo$_{0.8}$Ti$_{0.2}$N$_{x}$ films. For the fcc (Mo$_{0.8}$Ti$_{0.2}$)$_2$N$_{1.06}$ film, the low-temperature magnetoresistance isotherms cross at a single point and obey the power-law dynamical scaling. The $k_{\rm F}\ell$ of the (Mo$_{0.8}$Ti$_{0.2}$)$_2$N$_{1.06}$ film is the largest among all the films, which means the disorder strength of this film is the weakest. According to Vojta~\cite{jpaVojta}, the effects of quenched disorder on the critical point can be classified into three classes. (1) The average disorder strength decreases with increasing length scale, and the system tends to homogeneous at large length scales. As a result, the influence of the quenched disorder on the critical point becomes unimportant and can be neglected. (2) The system is inhomogeneous at all length scales, but the relative strength of inhomogeneities reaches to a finite value for large length scales. In this system, the critical point still exhibits conventional power-law scaling. However, the critical exponents become different from those of the clean system. (3) The relative strength of inhomogeneities increases with increasing length scale at any length scale. At these infinite-randomness critical points, the activated scaling instead of the power-law scaling is satisfied. As mentioned above, the fcc Mo$_2$N is a stable phase~\cite{prbKanoun} while the \emph{B}1-MoN is a metastable phase~\cite{jlcm}. Thus, the strength of the inhomogeneities in the fcc (Mo$_{0.8}$Ti$_{0.2}$)$_2$N$_{1.06}$ film could be asymptotically unimportant or approach a finite value for large length scale. As a result, the critical point maintains the power-law scaling on the whole.

\section{Conclusion}
A series of $\sim$80-nm-thick epitaxial MoTiN films with NaCl-type structure were grown on (100) MgO single crystal substrates by reactive sputtering method in an Ar and N$_2$ mixture atmosphere, and their low-temperature electrical transport properties were systematically studied. The composition of the $P_{\rm {N_2}}\simeq10.0\%$ film is (Mo$_{0.8}$Ti$_{0.2}$)$_2$N$_{1.06}$, while the compositions of the $P_{\rm {N_2}}\simeq12.5\%$, $P_{\rm {N_2}}\simeq15.0\%$, and $P_{\rm {N_2}}\simeq17.5\%$ films are Mo$_{0.8}$Ti$_{0.2}$N$_{0.84}$, Mo$_{0.8}$Ti$_{0.2}$N$_{1.05}$, and Mo$_{0.8}$Ti$_{0.2}$N$_{1.12}$, respectively. All the films reveal superconducting properties at low temperatures and are 3D with respect to superconductivity. A magnetic field, which is either perpendicular or parallel to the film plane, can drive the films transforming from superconducting to weakly disordered metal states. For the (Mo$_{0.8}$Ti$_{0.2}$)$_2$N$_{1.06}$ film, the low-temperature magnetoresistance isotherms cross at one single point and the resistivities near the SMT can be described by a power-law scaling deduced from the dirty boson model, which is applicable for both the perpendicular and parallel field. For the  Mo$_{0.8}$Ti$_{0.2}$N$_{x}$ ($0.84\lesssim x\lesssim 1.12$) films, both the critical exponents $z\nu_{\perp}$ and $z\nu_{\parallel}$ diverge as the systems approach the quantum critical points. In addition, both the perpendicular and parallel effective resistivities obey an activated scaling law deduced in framework of the random transverse-field Ising model. Our results provide strong experimental evidences for the existence of QGS in 3D superconductors.
\begin{acknowledgments}
The authors are grateful to Prof. Juhn-Jong Lin for valuable discussion. This work is supported by the National Natural Science Foundation of China through Grant No. 12174282.
\end{acknowledgments}


\begin{thebibliography}{56}
\bibitem{note3} We use $\tilde{B}_c^{\ast}$ here to distinguish the characteristic field $B_c^{\ast}$ in Eq.~(\ref{Eq-Bc}).
\end{thebibliography}
\end{document}